\documentclass[aps,preprint,prl,superscriptaddress]{revtex4}

\usepackage{graphicx}
\usepackage{amsmath}
\usepackage{amssymb}
\usepackage{units}
\usepackage{color}

\usepackage{setspace}
\usepackage[final]{changes}

\begin{document}

\title{Ultrafast Momentum Imaging of Pseudospin-Flip Excitations in Graphene}

\author{S. Aeschlimann}
\email{sven.aeschlimann@mpsd.mpg.de}
\author{R. Krause}
\author{M. Ch{\'a}vez-Cervantes}
\author{H. Bromberger}
\affiliation{Max Planck Institute for the Structure and Dynamics of Matter, Center for Free Electron Laser Science, Hamburg, Germany}
\author{R. Jago}
\author{E. Mali\'c}
\affiliation{Department of Physics, Chalmers University of Technology, Gothenburg, Sweden}
\author{A. Al-Temimy}
\author{C. Coletti}
\affiliation{Center for Nanotechnology @ NEST, Istituto Italiano di Tecnologia, Pisa, Italy}
\author{A. Cavalleri}
\affiliation{Max Planck Institute for the Structure and Dynamics of Matter, Center for Free Electron Laser Science, Hamburg, Germany}
\affiliation{Department of Physics, Clarendon Laboratory, University of Oxford, Oxford, United Kingdom}
\author{I. Gierz}
\email{isabella.gierz@mpsd.mpg.de}
\affiliation{Max Planck Institute for the Structure and Dynamics of Matter, Center for Free Electron Laser Science, Hamburg, Germany}

\date{\today}

\begin{abstract}

The pseudospin of Dirac electrons in graphene manifests itself in a peculiar momentum anisotropy for photo-excited electron-hole pairs. These interband excitations are in fact forbidden along the direction of the light polarization, and are maximum perpendicular to it. Here, we use time- and angle-resolved photoemission spectroscopy to investigate the resulting unconventional hot carrier dynamics, sampling carrier distributions as a function of energy and in-plane momentum. We first show that the rapidly-established quasi-thermal electron distribution initially exhibits an azimuth-dependent temperature, consistent with relaxation through collinear electron-electron scattering. Azimuthal thermalization is found to occur only at longer time delays, at a rate that depends on the substrate and the static doping level. Further, we observe pronounced differences in the electron and hole dynamics in n-doped samples. By simulating the Coulomb- and phonon-mediated carrier dynamics we are able to disentangle the influence of excitation fluence, screening, and doping, and develop a microscopic picture of the carrier dynamics in photo-excited graphene. Our results clarify new aspects of hot carrier dynamics that are unique to Dirac materials, with relevance for photo-control experiments and optoelectronic device applications.

\end{abstract}

\maketitle

The existence of anisotropic photo-carrier distributions in graphene was predicted \cite{TrushinEuroLett2011,MalicPRB2011} and observed in optical pump-probe experiments \cite{MittendorffNano2014, YanPRB2014, TrushinPRB2015, OttoPRL2016}, which showed a pronounced difference in the time-dependent optical response for different probe polarizations. The decay of the anisotropy extracted in this manner was attributed to optical phonon emission \cite{MalicPRB2011, MittendorffNano2014, YanPRB2014, TrushinPRB2015, MalicAPL2012, SatouJPhysConf2015}. However, a complete picture for these non-equilibrium phenomena can only be obtained by tracking both carrier energy and momentum in the time domain.

Here we use time- and angle-resolved photoemission spectroscopy (tr-ARPES) at extreme ultraviolet (XUV) wavelengths to track the temporal evolution of the photo-excited carrier distribution as a function of energy and momentum. We establish a hierarchy of events that redistribute carriers on the Dirac cone, including the formation of a quasi-thermal state with an azimuth-dependent anisotropic electron temperature, which indicates that primary thermalization occurs through collinear electron-electron scattering. Azimuthal relaxation through phonon emission and non-collinear electron-electron scattering plays a role only at later time delays, and is found to be strongly influenced by the substrate and the type of static doping of the graphene layer. Furthermore, the finite doping in our samples breaks the electron-hole symmetry and results in different dynamics for electrons and holes. Microscopic simulations of the anisotropic carrier dynamics indicate that the observed dynamics are due to a subtle interplay between doping that affects the scattering phase space and substrate screening which reduces the influence of electron-electron scattering.

Two different kinds of graphene samples were used for the present investigation. N-doped monolayer samples with an equilibrium chemical potential of $\mu_e=+0.4$\,eV and an effective screening constant of $\epsilon = 22$ \cite{WalterPRB2011} were obtained by thermal decomposition of the silicon face of SiC  \cite{VanBommelSurfSci1975, StarkeMRS2012}. P-doped samples with the chemical potential at $\mu_e=-0.2$\,eV and an effective screening constant of $\epsilon = 4.4$ \cite{WalterPRB2011} were instead obtained by decoupling the first inactive carbon monolayer formed by thermal decomposition of the same SiC face by hydrogen intercalation \cite{RiedlPRL2009, StarkeMRS2012}. After growth, these samples were exposed to air, characterized by Raman spectroscopy, reinserted into ultrahigh vacuum, and cleaned via annealing at 800$^\circ$C.

\begin{figure}
	\center
  \includegraphics[width = 0.5\columnwidth]{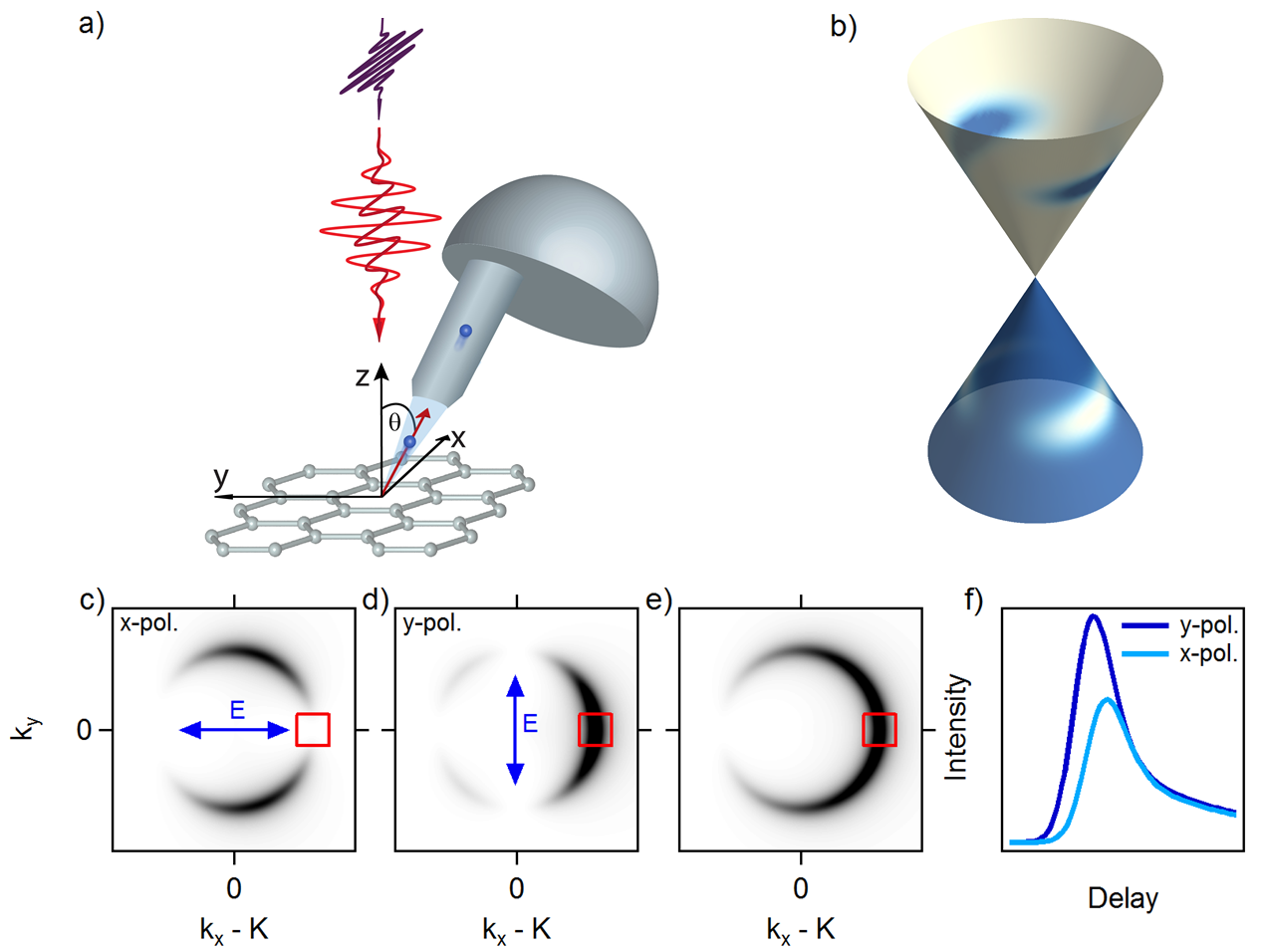}
  \caption{a) Sketch of the experimental setup. The sample is excited with x- or y-polarized pump pulses (red). Photoelectrons are ejected with x-polarized XUV probe pulses (violet) and pass through a hemispherical analyzer. b) Expected anisotropic charge carrier distribution after photo-excitation of monolayer graphene. Occupied and empty states are shown in blue and white, respectively. c)-e) Expected photoemission spectra at constant energy $E=E_D+\hbar\omega_{\text{pump}}/2$ as a function of $k_x$ and $k_y$ in the first instant after photo-excitation with x- (c) and y-polarized light (d) and the expected spectrum of an isotropic distribution (e). f) Sketch of the expected temporal evolution of the number of carriers inside the red box shown in (c), (d) and (e).}
  \label{fig1}
\end{figure}

The tr-XUV-ARPES experiments were performed at the MPISD in Hamburg. A Titanium:Sapphire amplifier operating at 1\,kHz repetition rate was used to generate synchronized 800nm optical pump and XUV probe pulses. The latter were obtained by high harmonic generation in an Argon gas jet. The 17th harmonic at $\hbar\omega_{\text{probe}}=26.3$\,eV was selected with a time-preserving grating monochromator \cite{FrassettoOptExpress2011} and used to measure photoelectron distributions from the sample. The probe polarization was fixed along the x axis (Fig. \ref{fig1}a). The polarization of the pump pulses was switched between x and y by rotating a half-wave plate. Both pump and probe impinged onto the sample at normal incidence. The experimental data shown in this work was obtained with pump fluences ranging from 1.3 to 2.8\,mJ/cm$^2$. The energy and temporal resolution of the tr-ARPES experiment were 350\,meV and 145\,fs, respectively.

For the experiments reported here, we used a hemispherical analyzer with the entrance slit parallel to the x axis, to measure the photocurrent as a function of energy and in-plane momentum $k_x$ (Fig. \ref{fig1}a). In order to record the complete Dirac cone (photocurrent as a function of $k_x$, $k_y$, and energy) we rotated the sample around the x axis.

Pump pulses at $\hbar\omega_{\text{pump}}=1.5$\,eV generated electron-hole pairs at $E_D \pm \hbar\omega_{\text{pump}}/2$, where $E_D$ is the energy of the Dirac point where conduction and valence band meet (Fig. \ref{fig1}b). This process mapped valence band states onto conduction band states of opposite pseudospin. Hence, optical excitation involved pseudospin flips which resulted in an angle-dependent transition probability $|M_{\text{pump}}|^2 \propto \sin^2(\phi_k - \phi_A^{\text{pump}})$ \cite{TrushinEuroLett2011, MalicPRB2011}, where $\phi_k$ and $\phi_A^{\text{pump}}$ are the angles between the k-vector of the electron or the pump polarization and the x axis, respectively. As immediately evident from the expression above, the transition probability was then zero along the direction of the electric field ($\phi_k = \phi_A^{\text{pump}}$) and maximum perpendicular to it. 

Note also that the photocurrent is subject to momentum-dependent matrix element effects. The photoemission cross section in graphene is proportional to $|M_{\text{probe}}|^2\propto 1/2 (1 \pm \cos(\phi_k - 2\phi_A^{\text{probe}}))$ \cite{ShirleyPRB1995, DaimonJSRP1995, BostwickNJP2007} with the upper (lower) sign for the conduction (valence) band and $\phi_A^{\text{probe}}=0$ in the present experiment, which turns part of the Dirac cone invisible. The photoelectron distribution can then be obtained by multiplying the actual carrier distribution with $|M_{\text{probe}}|^2$.

Figures \ref{fig1}c-e illustrate the expected photoelectron spectrum at $E_D + \hbar\omega_{\text{pump}}/2$ as a function of $k_x$ and $k_y$ for excitation with x- and y-polarized light, and, for comparison, for a homogeneous carrier distribution. Figure \ref{fig1}f shows the expected evolution in time of the photocurrent inside the red box in Figs. \ref{fig1}c-e \cite{MalicPRB2011, MalicAPL2012, SatouJPhysConf2015, MittendorffNano2014, YanPRB2014, TrushinPRB2015, OttoPRL2016}. For pump pulses polarized along the x axis, the carriers are expected to fill these states only after scattering around the cone. Hence, we expect to measure a delayed rise and a lower peak signal for excitation with x-polarized light compared to excitation with y-polarized light. We also expect the two curves to overlap once the distribution becomes isotropic, before further cooling by optical and acoustic phonon emission occurs at longer time delays \cite{KampfrathPRL2005, YanPRB2009, KangPRB2010, WinzerNanoLett2010, BreusingPRB2011, SongPRL2012, GrahamNatPhys2013, BetzNatPhys2013, JohannsenPRL2013}.

\begin{figure}
	\center
  \includegraphics[width = 1\columnwidth]{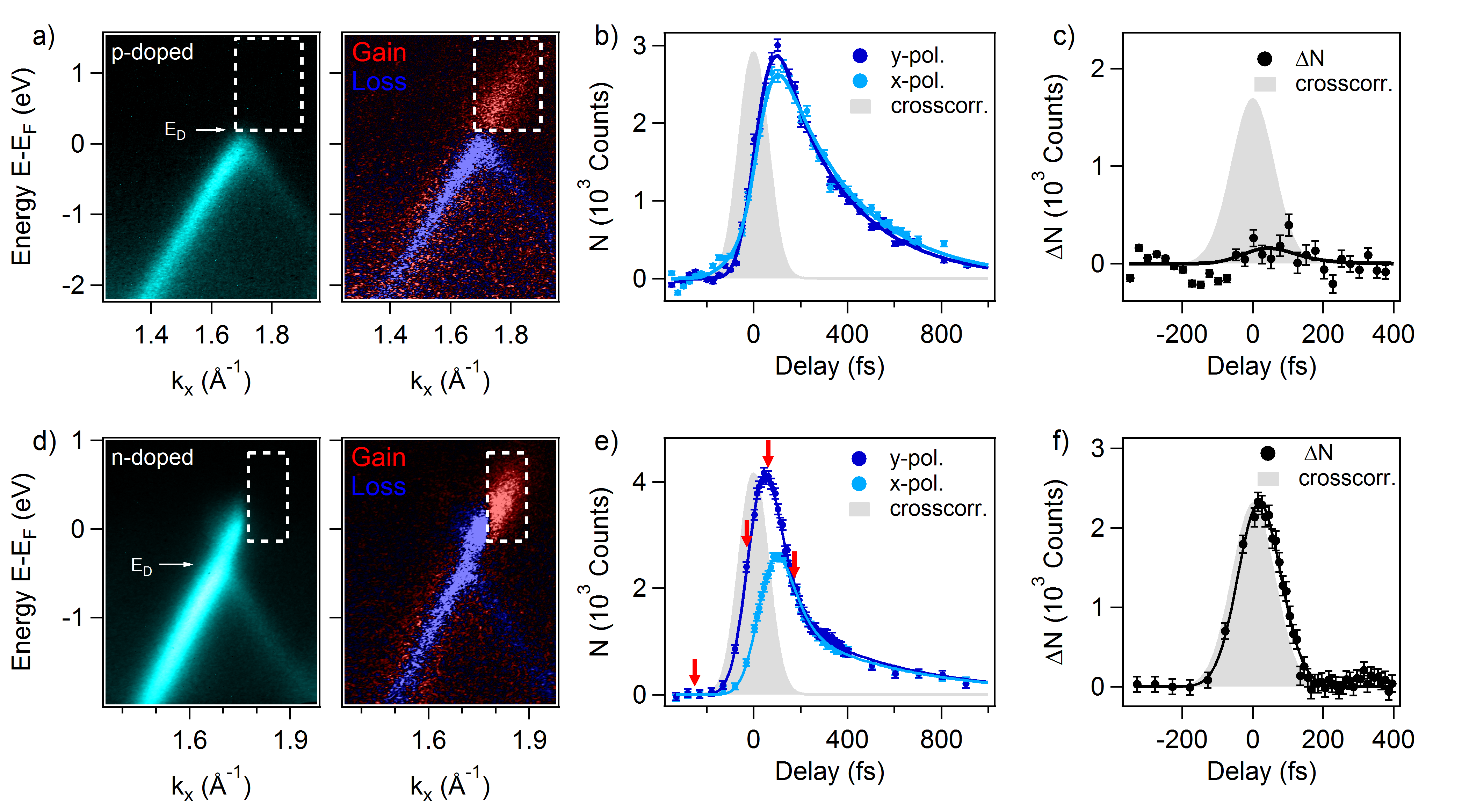}
  \caption{Photoemission data for p-doped (upper panel, excitation fluence of 1.5\,mJ/cm$^2$) and n-doped graphene (lower panel, excitation fluence of 2.8\,mJ/cm$^2$): a), d) ARPES spectra for negative time delays and pump-induced changes of the photocurrent for y-polarized pump pulses at the peak of the pump-probe signal. b), e) photocurrent integrated over the area of the white boxes in (a) and (d) versus pump-probe delay for x- (light blue) and y-polarized pump pulses (dark blue). The respective difference in intensity is shown in (c) and (f). The light gray area represents the temporal cross-correlation of pump and probe pulses. Tr-ARPES data for the n-doped sample for an excitation fluence of 1.3\,mJ/cm$^2$ is shown in \cite{SupMat}.}
  \label{fig2}
\end{figure}

In a first set of experiments we measured the photocurrent as a function of energy and $k_x$, and compared the effect of x- and y-polarized excitation in p- and n-doped samples (upper and lower panel of Fig. \ref{fig2}, respectively). Figures \ref{fig2}a and d show ARPES snapshots at a negative pump-probe delay and pump-induced changes of the photocurrent at the pump-probe delay at which the signal was maximum.
In order to compare the number of excited carriers for x- and y-polarized pump pulses we integrated the photocurrent over the area indicated in Figs. \ref{fig2}a and d (white boxes). The time-dependent photocurrent is shown in Figs.  \ref{fig2}b and e. These data were fitted with an error function and a double exponential decay. We also show the temporal cross-correlation between pump and probe pulses (gray-shaded area), as obtained from the temporal derivative of the error function, with a full width at half maximum of 145\,fs. For p-doped samples, the pump-probe signal for x- and y-polarized pump pulses was found to be the same within the error bars. On the contrary, we found a pronounced difference between the two pump polarizations for the n-doped sample, indicating the presence of a long-lived anisotropic carrier distribution. In Figs. \ref{fig2}c and f we plot the time-dependent anisotropy (difference between the dark and light blue curves in Figs. \ref{fig2}b and e), which was found to relax at a rate limited by the time resolution of the experiment.

\begin{figure}
	\center
  \includegraphics[width = 1\columnwidth]{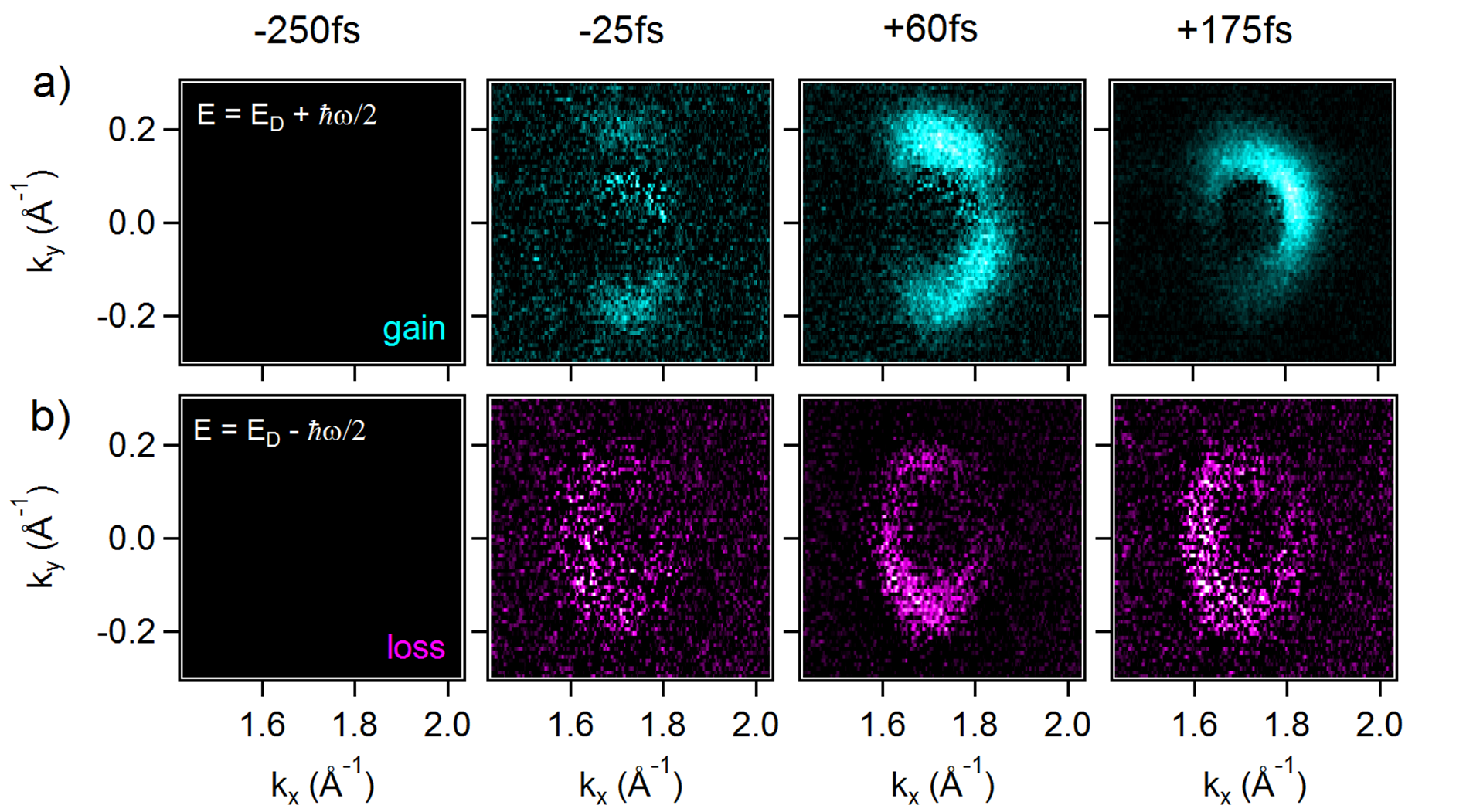}
  \caption{Photoemission spectra at constant energy $E=E_D+\hbar\omega_{\text{pump}}/2$ (panel a) and $E=E_D-\hbar\omega_{\text{pump}}/2$ (panel b) as a function of $k_x$ and $k_y$ for an excitation fluence of 2.8\,mJ/cm$^2$ at four different time delays as indicated by red arrows in Fig. \ref{fig2}e. Note that the sickle-shaped image at $\text{t}=175$\,fs is slightly rotated away from the $k_x$ axis due to a small azimuthal misalignment of the sample and that the photoemission cross section for the valence band leads to zero intensity on the opposite side of the Dirac cone compared to the conduction band.}
  \label{fig3}
\end{figure}

Time-dependent carrier distributions for all $k_x$ and $k_y$ values were measured for n-doped samples and x-polarized pump pulses. Constant-energy cuts integrated over an interval of $\pm$50\,meV around $E_D+\hbar\omega_{\text{pump}}/2$ are reported for four different delays (Fig. \ref{fig3}a), indicated by red arrows in Fig. \ref{fig2}e. At negative delay ($\text{t}=-250$\,fs) no excited carriers are detected. For a time delay of $\text{t}=-25$\,fs, that is half way through the rising edge, the anisotropic carrier distribution is already observable, reaching its maximum at $\text{t}=+60$\,fs. At $\text{t}=+175$\,fs the carrier distribution becomes isotropic, with an angular dependence caused by the photoemission matrix element alone. The measured spectra nicely agree with the expectations shown in Figs. \ref{fig1}c-e. For comparison, we also show the photo-excited hole distribution at $E_D-\hbar\omega_{\text{pump}}/2$ in Fig. \ref{fig3}b. \added{Note that the photoemission cross section for the valence band is flipped with respect to the one of the conduction band. Sketches of the expected measured hole distribution can be obtained by mirroring Fig. \ref{fig1}c-e on the $k_y$-axis. The measured hole distribution (Fig. \ref{fig3}b) shows a much smaller anisotropy than the measured electron distribution (Fig. \ref{fig3}a). A more detailed comparison between electron and hole dynamics is given in the supplementary material \cite{SupMat}.}

\begin{figure}
	\center
  \includegraphics[width = 0.5\columnwidth]{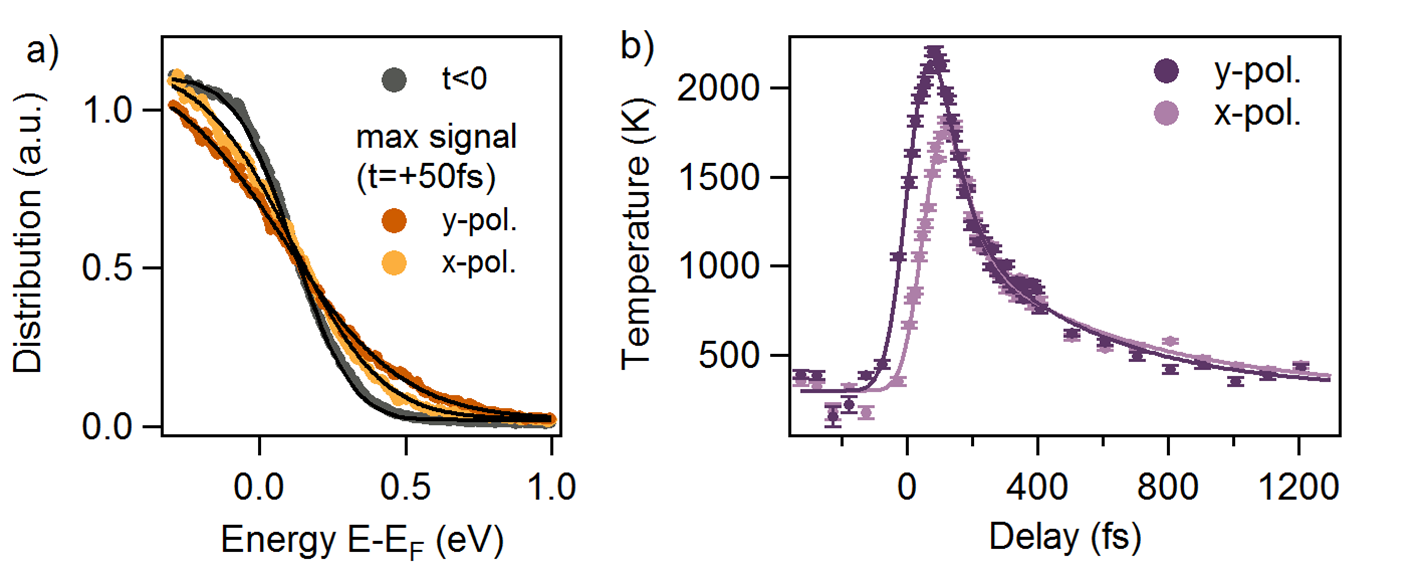}
  \caption{a) Electron distribution functions along the $k_x$ direction for n-doped graphene. Gray curves show the distribution at negative pump-probe delay, light and dark orange curves show the respective distributions at $\text{t}=50$\,fs for x- and y-polarized pump pulses. Black curves are Fermi-Dirac fits. b) Temporal evolution of the electron temperature obtained from the fits in (a).}
  \label{fig4}
\end{figure}

By integrating the two-dimensional ARPES spectra in Fig. \ref{fig2}d along $k_x$ for x- and y-polarized pump pulses, we obtained transient electron distribution functions \cite{GierzNatureMat2013, UlstrupRevSciInstr2014} at the minima and maxima of $|M_{\text{pump}}|^2$, respectively, along the direction where the photoemission cross section is maximum. The gray data points in Fig. \ref{fig4}a show the distribution at negative delay. Light and dark orange data points show the distributions for x- and y-polarized pump pulses at $\text{t}=+50$\,fs where the pump-probe signal reaches its maximum for excitation with y-polarized light. The black lines are Fermi-Dirac fits convolved with a Gaussian with a full width at half maximum of 350\,meV to account for the finite energy resolution. The temporal evolution of the resulting electron temperature is shown in Fig. \ref{fig4}b. At early times, the electron temperature along $k_x$ is found to be smaller for x-polarized pump pulses than for y-polarized pump pulses.


We first note that the electron distribution can be described with a Fermi-Dirac distribution at all pump-probe delays (Fig. \ref{fig4}a), indicating that electron-electron scattering thermalizes the photo-excited carriers on a time scale short compared to our temporal resolution. The observed pump-polarization dependence of the electron temperature (Fig. \ref{fig4}b) shows that this transient quasi-thermal state has an azimuth-dependent temperature and provides direct evidence that electron-electron scattering is strongly confined to lines pointing radially away from the Dirac point as predicted in \cite{MalicPRB2011, MalicAPL2012}.

Relaxation around the cone, which re-establishes an isotropic carrier distribution, can in principle occur through electron-phonon scattering or non-collinear electron-electron scattering. While the decay of the anisotropy is believed to be dominated by phonon emission in the low fluence regime \cite{MalicPRB2011,MittendorffNano2014,MalicAPL2012}, we expect non-collinear electron-electron scattering to be of similar importance for the high excitation fluences applied in this work. In order to develop a microscopic understanding of the scattering channels that are responsible for the decay of the anisotropy in the present study, we simulate the influence of pump fluence, substrate screening, and doping on the anisotropic carrier dynamics in graphene. Details are given in the Supplemental Material \cite{SupMat}. In Fig. \ref{fig5} we present the simulated dynamics of the anisotropy for the two different graphene samples for a pump fluence of 1.5\,mJ/cm$^2$. In agreement with the experiment we find a larger and longer-lived anisotropy for the n-doped sample (Fig. \ref{fig5}a) compared to the p-doped sample (Fig. \ref{fig5}b). The reason for the enhanced lifetime of the anisotropy in the n-doped sample can be traced back to the large value of the chemical potential that reduces the scattering phase space for both electron-electron (dotted lines in Fig. \ref{fig5}a and b) and electron-phonon scattering (dashed lines in Fig. \ref{fig5}a and b) as well as the strong effective screening of the Coulomb interaction due to the large dielectric constant of the substrate. As the measured lifetime of the anisotropy in the present work is resolution limited, the difference in lifetime shows up as a difference in amplitude of the measured anisotropy. Our microscopic simulations are also able to reproduce the difference between electron and hole dynamics (see Supplemental Material \cite{SupMat}). This can be explained by the finite positive value of the chemical potential that breaks the electron-hole symmetry and increases (decreases) the scattering phase space for holes (electrons).

In summary, we have used time- and angle-resolved photoemission spectroscopy to visualize anisotropic photo-carrier distributions in p- and n-doped monolayer graphene. We found that collinear electron-electron scattering rapidly thermalizes the carriers along lines pointing radially away from the Dirac point, leading to a quasi-thermal state with an azimuth-dependent electron temperature. We also observed that the magnitude and the decay of the measured anisotropy are influenced by the underlying substrate and the doping level of the graphene layer and are different for electrons and holes. Using microscopic simulations of the anisotropic carrier dynamics we are able to explain the experimental observations by a subtle interplay of doping that modifies the scattering phase space and screening that reduces the efficiency of electron-electron scattering. Our results visualize photo-carrier dynamics that are unique to Dirac materials, in which the pseudospin is responsible for peculiar anisotropic photo-carrier distributions. We also note that the ability to tune hot carrier dynamics via doping or screening might potentially be exploited in graphene-based thermoelectric devices \cite{MuellerNatPhot2010, BonaccorsoNatPhot2010, GaborScience2011, SunNatNano2012, EchtermeyerNano2014}, or other opto-electronic applications of this class of solids.

This work received financial support from the German Research Foundation through the Priority Program SPP 1459 and the Collaborative Research Center SFB 925 as well as the European Union's Horizon 2020 Research and Innovation Programme under Grant Agreement No. 696656-GrapheneCore1.

\begin{figure}
	\center
  \includegraphics[width = 0.5\columnwidth]{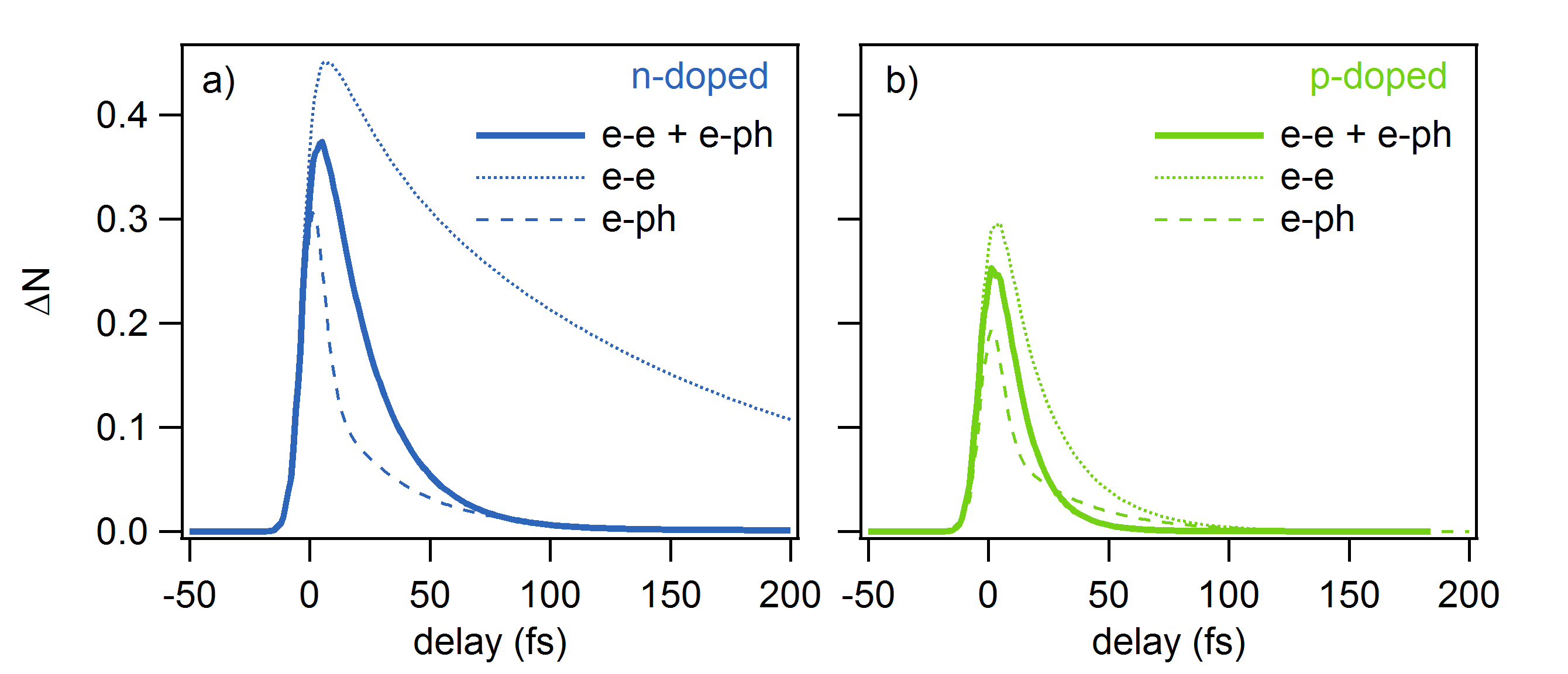}
  \caption{Simulated dynamics of the anisotropy in the n-doped (a) and the p-doped sample (b). Solid lines represent the full dynamics, dotted and dashed lines represent the dynamics for electron-electron and electron-phonon scattering only, respectively.}
  \label{fig5}
\end{figure}

\clearpage

\section{Supplementary Material}
\label{sec:Supplementary}

\section{additional tr-ARPES data}

\begin{figure*}
 \includegraphics[width=\linewidth]{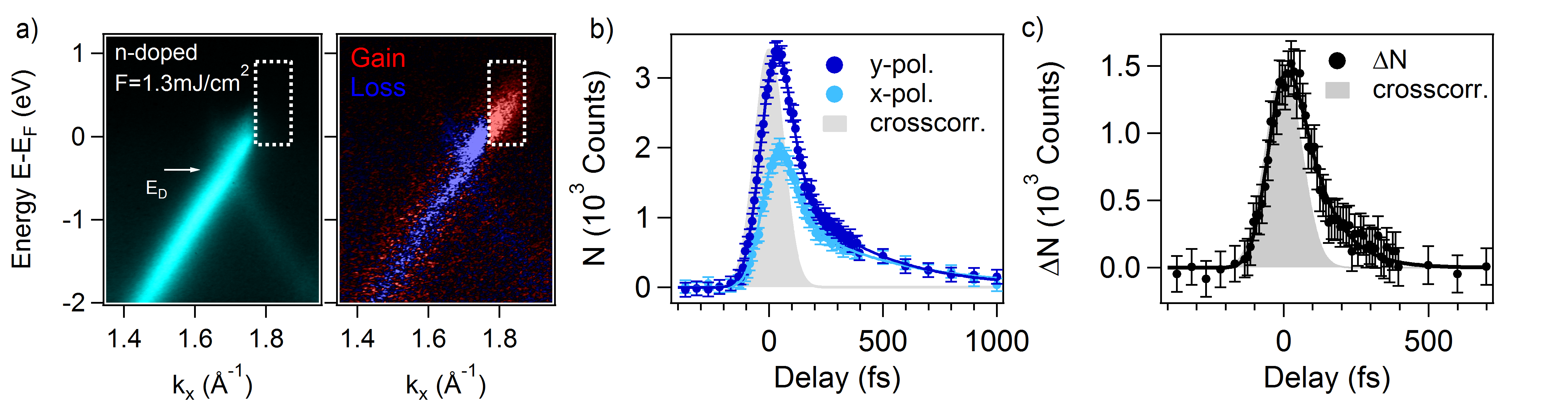}
\caption{\textbf{Low-fluence tr-ARPES data for n-doped sample.} (a) Snapshot of the band structure at negative pump-probe delay together with pump-induced changes of the photocurrent at the time delay corresponding to the peak of the pump-probe signal for excitation with y-polarized light. (b) Intensity integrated over the area indicated by the white box in panel a for x- (light blue) and y-polarized pump pulses (dark blue) as a function of pump-probe delay. (c) Anisotropy $\Delta N$ given by the difference between the two curves in panel b. The grey-shaded area represents the temporal crosscorrelation between infrared pump and ultraviolet probe pulse.}
\label{fig_data}
\end{figure*} 

In Fig. 2 of the main text we provide data for the p-doped sample at an excitation fluence of 1.5\,mJ/cm$^2$ and for the n-doped sample at an excitation fluence of 2.8\,mJ/cm$^2$. As discussed in detail below we expect the anisotropy to be fluence dependent. For a proper comparison between the two samples an excitation with the same fluence is desirable. In Fig. \ref{fig_data} we provide tr-ARPES data for the n-doped sample at an excitation fluence of 1.3\,mJ/cm$^2$, similar to the one for the p-doped sample in the main text. The reason why this data is not reported in the main text, is that in this measurement there is a time-zero drift between the data recorded with x- and y-polarized pump pulses, respectively, due to unstable air conditioning in the laboratory on that particular day. Due to this time-zero drift the anisotropy $\Delta N$ (obtained by subtracting the two data sets recorded with x- and y-polarized pump pulses for a given time delay, see Fig. \ref{fig_data}b and c) cannot be determined properly. We want to stress that the difference between the two peak intensities in Fig. \ref{fig_data}b is similar to the one in Fig. 2e of the main text where we show the tr-ARPES data for the n-doped sample for a higher pump fluence of 2.8\,mJ/cm$^2$. This indicates that increasing the pump fluence from 1.3 to 2.8\,mJ/cm$^2$ has a negligible effect on the amplitude of the measured anisotropy. This justifies the comparison made in Fig. 2 of the main text.

\begin{figure*}
 \includegraphics[width=\linewidth]{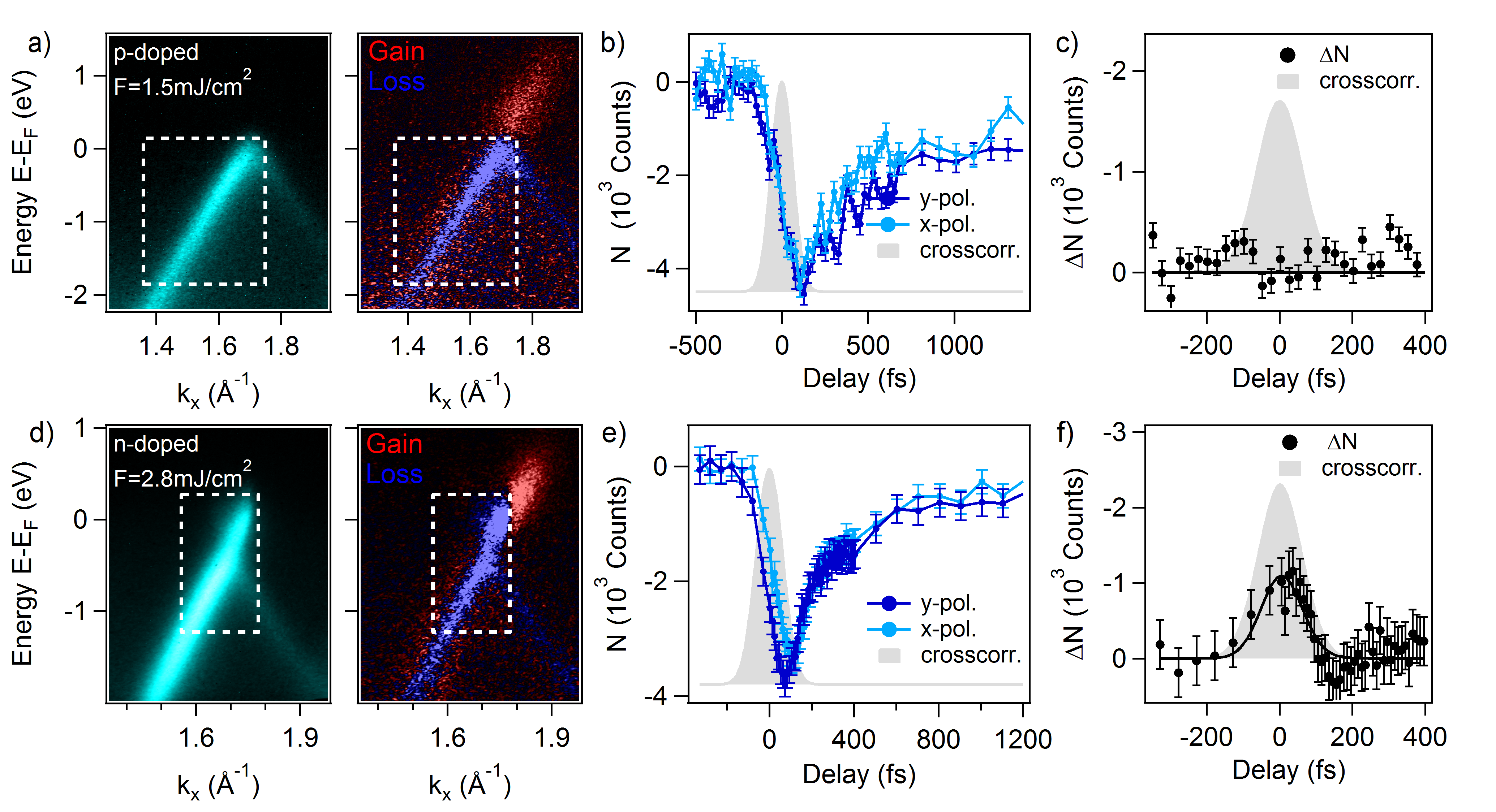}
\caption{\added{\textbf{Hole Dynamics.} Upper pannel: p-doped graphene, excitation fluence of 1.5\,mJ/cm$^2$, lower panel: n-doped graphene, excitation fluence of 2.8\,mJ/cm$^2$. a), d) ARPES spectra for negative time delays and pump-induced changes of the photocurrent for y-polarized pump pulses at the peak of the pump-probe signal. b), e) change of photocurrent integrated over the area of the white boxes in (a) and (d) versus pump-probe delay for x- (light blue) and y-polarized pump pulses (dark blue). The respective difference in intensity is shown in (c) and (f). The light gray area represents the temporal cross-correlation of pump and probe pulses.}}
\label{fig_hole_dynamics}
\end{figure*}

\added{In Fig. \ref{fig_hole_dynamics} we present the hole dynamics in p- and n-doped samples in analogy to the electron dynamics in Fig. 2 in the main text. We find that the anisotropy of the holes is less pronounced than the one of the electrons for the n-doped sample. In the p-doped sample, we don't observe any anisotropy for electrons or holes.}

\section{microscopic simulations}

Here, we present the theoretical study of the anisotropic carrier dynamics in graphene. The applied microscopic many-particle approach allows us to resolve the dynamics of optically excited electrons in time, energy, and momentum with respect to the polarization of the excitation pulse \cite{malic13, MalicPRB2011, malic16}. 
The core of the approach is built by Graphene Bloch Equations (GBE) that have been derived in the density matrix formalism within the second-order Born-Markov approximation \cite{lindberg88, haug04, rossi02, knorr96, kira06, malic13}. We obtain a system of coupled equations of motion for the carrier occupation $\rho^{\lambda}_{\bf k}=\langle a_{\lambda\bf k}^{+}a^{\phantom{+}}_{\lambda\bf k}\rangle$ in the state $(\bf{k}, \lambda)$ characterized by the momentum $\bf k$ and the band index $\lambda$, the microscopic polarization $p_{\bf k}=\langle a_{v\bf k}^{+}a^{\phantom{+}}_{c\bf k}\rangle $ that is a measure for the transition probability between the valence ($\lambda=v$) and the conduction ($\lambda=c$) band, and phonon number $n_{\bf q}^j=\langle b_{j\bf q}^{+}b^{\phantom{+}}_{j\bf q}\rangle$ in the considered optical or acoustic phonon mode $j$ with the phonon momentum $\bf q$. 
Here, we have expressed the microscopic quantities in the formalism of second quantization introducing creation and annihilation operators for electrons ($a_{{\lambda\bf k}}^{+}, a_{{\lambda\bf k}}^{\phantom{+}}$) and phonons ($b_{{j\bf q}}^{+}, b_{{j\bf q}}^{\phantom{+}}$) \cite{kadi14}.

Applying the Heisenberg equation of motion and exploiting the fundamental commutation relations for fermions and bosons, we derive the GBE in second-order Born-Markov approximation yielding \cite{malic13}
\begin{eqnarray}
\label{bloch_p}
 \dot{p}_{\bf k}&=&i\Delta\omega_{\bf k}p_{\bf k}-i\Omega_{\bf k}^{vc}\big(\rho_{\bf k}^c-\rho_{\bf k}^v\big)+\mathcal{U}_{\bf k}-\gamma_{\bf k} p_{\bf k},\\[8pt]
\label{bloch_rho}
 \dot{\rho}_{\bf k}^{\lambda}&=&\pm2{\rm{Im}}\big(\Omega_{\bf k}^{vc*}p_{\bf k}\big)+\Gamma_{\lambda,{\bf k}}^{\rm{in}}\big(1+\rho_{\bf k}^\lambda\big)-\Gamma_{\lambda,{\bf k}}^{\rm{out}}\rho_{\bf k}^\lambda,\\[8pt]
\label{bloch_nph}
\dot{n}_{\bf q}^j&=&\Gamma_{j,{\bf q}}^{\rm{em}}\big(1+n_{\bf q}^j\big)-\Gamma_{j,{\bf q}}^{\rm{abs}}n_{\bf q}^j-\gamma_j\big(n_{\bf q}^j-n_B\big)
\end{eqnarray}
with the energy difference $\hbar\Delta\omega_{\bf k}=(E_{\bf
  k}^{v}-E_{\bf k}^{c})$,  the Bose-Einstein distribution $n_B$ denoting the equilibrium distribution of phonons,  the phonon lifetime $\gamma^{-1}_j$, and the Rabi frequency
$\Omega^{vc}_{\bf k}(t)=i\frac{e_{0}}{m_{0}} {\bf{M}}_{{\bf k}}^{vc}\cdot {\bf{A}}(t)$ describing the optical excitation of graphene with ${\bf{M}}_{{\bf k}}^{vc}$ as the optical matrix element \cite{grueneis03, malic06}. All quantities in GBE (except for $n_B$ and $\Delta\omega_{\bf k}$) depend on time. The appearing matrix elements are calculated with tight-binding wave functions in the nearest-neighbor approximation \cite{malic13}, which is known to be a good approximation for graphene close to the Dirac point \cite{reich02b}. The doping is included via initial occupation of electrons and breaks the symmetry of the electron and hole dynamics \cite{kadi15}.
 The bare Coulomb potential $V({\bf q})$ appearing in the Coulomb-induced scattering rates $\Gamma_{\lambda,{\bf k}}^{\rm{in/out}}(t)$
 is screened due to the presence of many electrons and the substrate. The effects arising from the electrons in the core states and the surrounding subtrate are taken into account by introducing a dielectric background constant $\varepsilon_{bg}$. The screening stemming from other valence electrons are calculated within an effective single-particle Hamiltonian approach leading to the Lindhard approximation of the dielectric function $\varepsilon({\bf q})$ \cite{haug04,malic13}.
Since this many-particle-induced screening is directly influenced by carrier occupations in the conduction and valence bands, doping plays a crucial role and has a significant influence on the ultrafast
carrier dynamics in graphene.

The equations take into account all relevant two-particle relaxation channels including Coulomb- and phonon-induced intra- and interband as well as intra- and intervalley scattering processes. The time- and momentum-dependent scattering rates $\Gamma_{\lambda{\bf k}}(t)=\Gamma_{\lambda{\bf k}}^{\rm{c-c}}(t)+\Gamma_{\lambda{\bf k}}^{\rm{c-ph}}(t)$ entering the equation for the carrier occupation $\rho_{\bf k}^\lambda (t)$ describe the strength of the carrier-carrier (c-c) and carrier-phonon (c-ph) scattering processes. For the phonon occupation, we obtain the corresponding emission and absorption rates $\Gamma_{j,{\bf q}}^{\rm{em/abs}}$ \cite{malic11}. The many-particle scattering also contributes to  diagonal ($\gamma_{\bf k}(t)$) and off-diagonal dephasing ($\mathcal{U}_{\bf k}(t)$) of the microscopic polarization $p_{\bf k}(t)$.

More details on the theoretical approach can be found in Refs. \cite{MalicPRB2011, malic13}. 

The aim of the current study is to model the measured anisotropy in tr-ARPES experiments on p- and n-doped graphene samples in the strong excitation regime. We investigate the anisotropy defined as
\begin{equation}
 \Delta N=\frac{1}{\Delta E}\int_{E_F}^{E_{\rm{max}}}\left(\rho^{k_x}_E(t)- \rho^{k_y}_E(t)\right)\,dE,
\label{anis_av}
\end{equation}
where $\rho^{k_x}_E(t)$ and $\rho^{k_y}_E(t)$ are the time-resolved carrier occupations evaluated in the direction ($k_x$) and opposite to the direction ($k_y$) of the polarization of the pump pulse. We integrate from the Fermi energy $E_F$ up to a maximal energy well beyond the excitation energy to capture all of the photo-excited electrons. To get profound insights into the elementary processes behind the anisotropy, we investigate the impact of pump fluence, doping, and substrate screening on the anisotropy.

\subsection{Impact of pump fluence}

\begin{figure}
 \includegraphics[width=\linewidth]{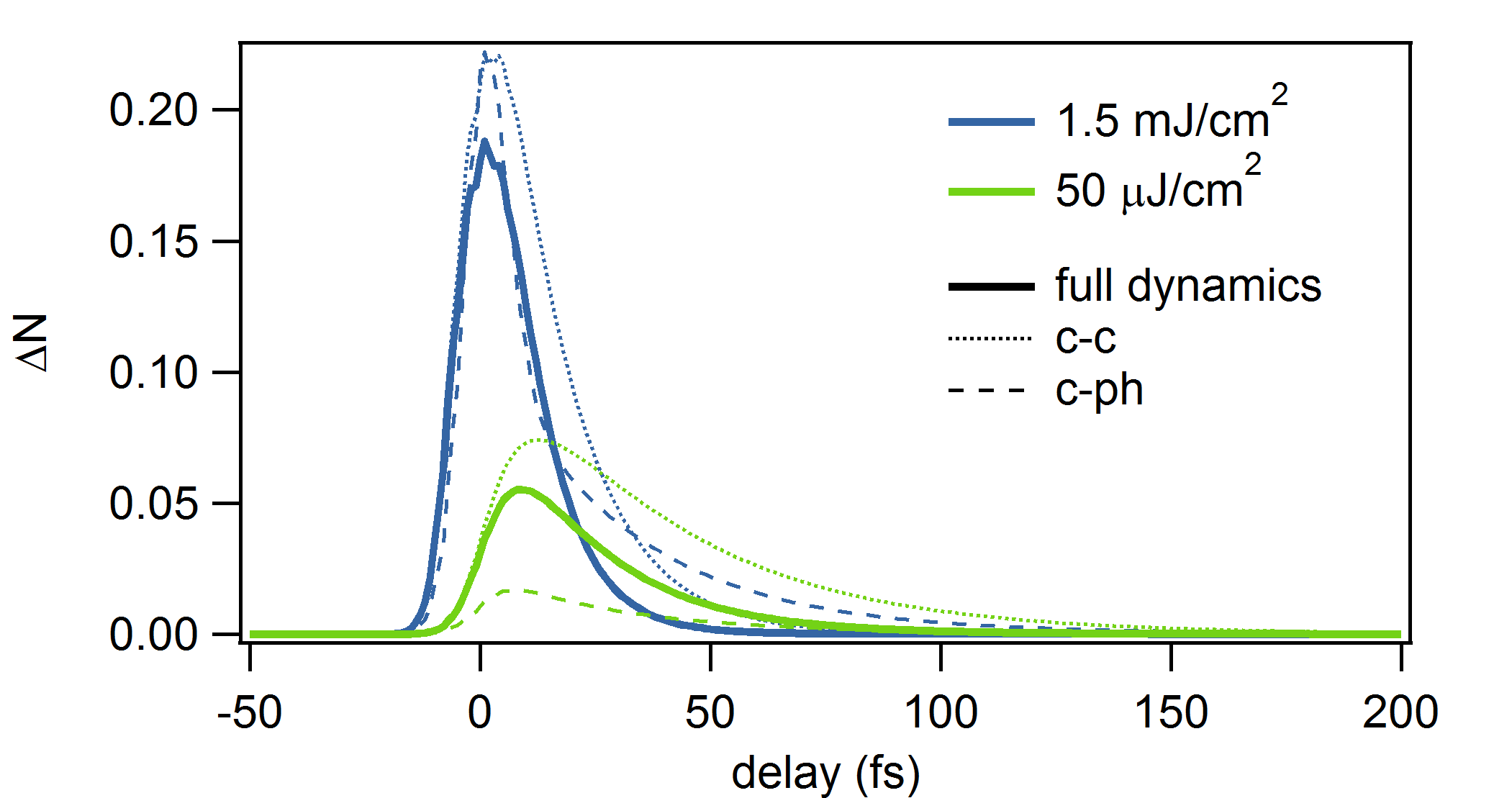}
\caption{\textbf{Fluence dependence.} Temporal evolution of the carrier anisotropy $\Delta N$ (see Eq. \ref{anis_av}) in an undoped graphene sample on SiC ($\varepsilon_s=7.8$) for a relatively small ($\unit[50]{\rm{\mu}J/cm^2}$, green) and a relatively large pump fluence ($\unit[1.5]{mJ/cm^2}$, blue). To reveal the role of single many-particle processes, we show the full dynamics (continuous lines) and compare it with the dynamics taking into account only carrier-phonon scattering (dashed lines) and carrier-carrier scattering (dotted lines).}
\label{fig_fluence}
\end{figure}

First, we focus on the dependence of the anisotropy on the applied pump fluence. We consider an undoped graphene sample on a SiC substrate with a dielectric constant of $\varepsilon_s=7.8$ and directly compare the temporal evolution of the anisotropy for two different pump fluences ($\unit[50]{\rm{\mu}J/cm^2}$ and $\unit[1.5]{mJ/cm^2}$, respectively) in Fig. \ref{fig_fluence}. The large pump fluence is comparable to the one used in the tr-ARPES measurements in the main text. To reveal the elementary processes responsible for the decay of the anisotropy, we show the full dynamics (continuous lines) as well as calculations taking into account only carrier-phonon (c-ph, dashed lines) or carrier-carrier scattering (c-c, dotted lines). The most obvious impact of the fluence is a significantly increased anisotropy $\Delta N$ (see Eq. \ref{anis_av}) in the strong excitation regime, simply reflecting the higher number of photo-excited carriers. Note that previous optical pump-probe measurements used a different definition for the anisotropy based on the pump-induced changes of the transmission $\Delta T$. In this case, the ratio between $\Delta T$ for parallel and cross-polarized pump and probe beams was found to decrease with increasing fluence \cite{TrushinPRB2015}. By comparing the dashed and dotted green lines in Fig. \ref{fig_doping}, we conclude that the decay of the anisotropy in the small fluence regime is dominated by c-ph scattering in good agreement with literature \cite{MalicPRB2011, MalicAPL2012, MittendorffNano2014}. This is because optically excited carriers can efficiently scatter across the Dirac cone via different phonon modes. In contrast, Coulomb interaction prefers collinear scattering events \cite{MalicAPL2012, MittendorffNano2014, OttoPRL2016}. The situation is different in the high fluence regime, where c-c becomes more efficient due to the large number of photo-excited electrons. In this case, non-collinear c-c scattering and c-ph scattering become of similar importance for the decay of the anisotropy (see the dotted and dashed blue lines in Fig. \ref{fig_fluence}). However, c-ph scattering still remains the dominant channel for the reduction of the anistropy in the first tens of femtoseconds. Non-collinear c-c scattering becomes important at later times, where the dotted and dashed blue lines in Fig. \ref{fig_fluence} cross. This crossover happens once the photo-excited electrons, after having emitted a couple of phonons, are too close to the Fermi level to allow for further phonon emission.

\subsection{Impact of doping}

\begin{figure}[t!]
 \includegraphics[width=\linewidth]{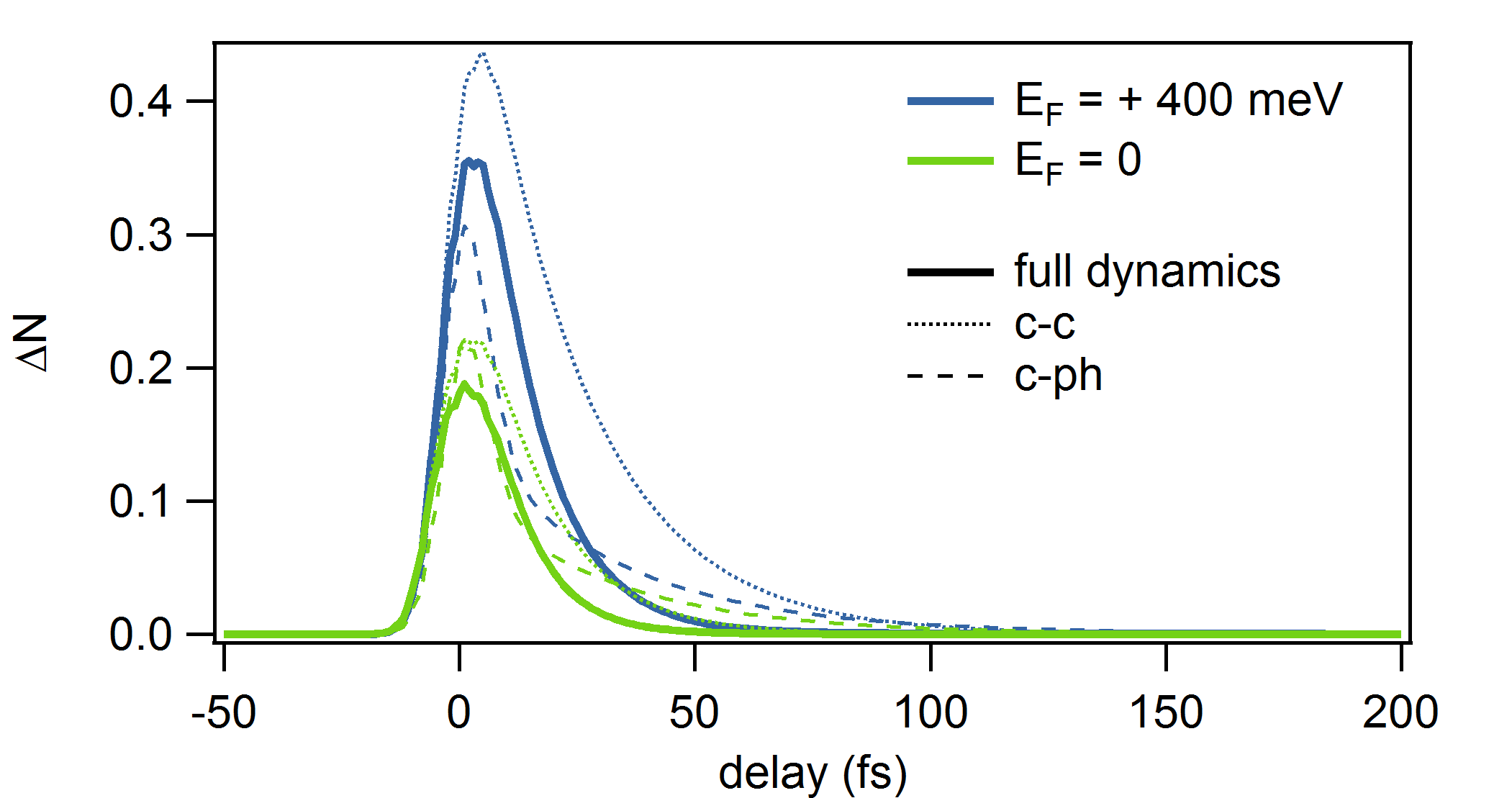}
\caption{\textbf{Doping dependence.} Temporal evolution of the anisotropy $\Delta N$ in the strong excitation regime ($\unit[1.5]{mJ/cm^2}$) for n- ($E_F$=\unit[400]{meV}, blue) and undoped graphene (green) on SiC ($\varepsilon_s=7.8$). The impact of Coulomb- (dotted line) and phonon-induced (dashed line) scattering channels is illustrated separately. The complete dynamics is shown as continuous lines.}
\label{fig_doping}
\end{figure}

Now we study the influence of doping for a given fluence ($\unit[1.5]{mJ/cm^2}$) and substrate screening ($\varepsilon_s=7.8$). We assume an n-doping of \unit[400]{meV}, the same as in the tr-ARPES experiments. We find a bigger amplitude of the anisotropy in the n-doped graphene sample compared to the undoped graphene sample (blue and green in Fig. \ref{fig_doping}, respectively) for the full dynamics as well as for carrier-carrier and carrier-phonon scattering only. We attribute this to a combination of a reduced scattering phase space for c-c and c-ph scattering as well as a bigger screening of the Coulomb interaction due to the increased number of free carriers in the n-doped graphene sample.

\subsection{Impact of substrate screening}

\begin{figure}[b!]
 \includegraphics[width=\linewidth]{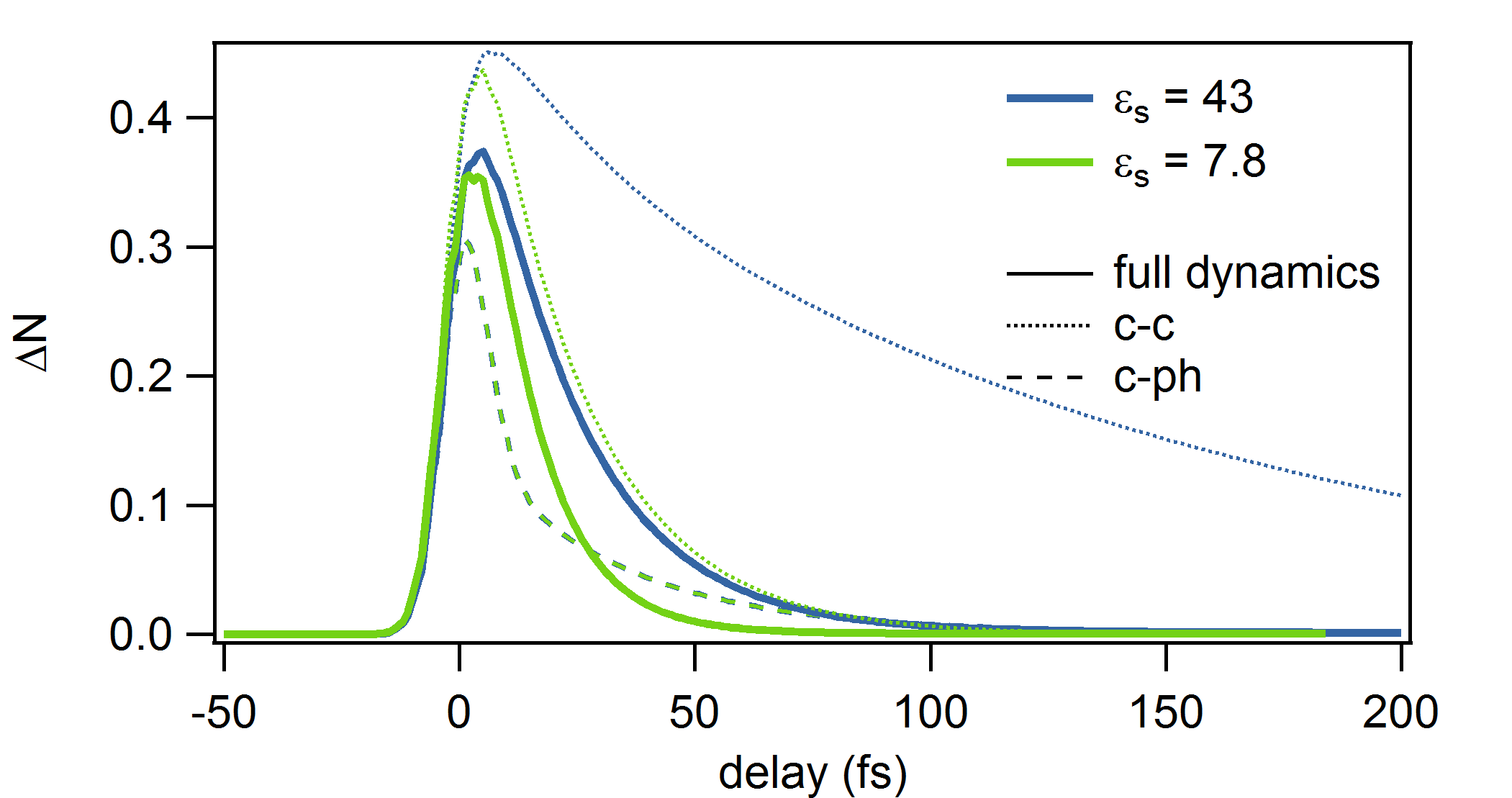}
\caption{\textbf{Substrate dependence.} Temporal evolution of the carrier anisotropy $\Delta N$ in the strong excitation regime ($\unit[1.5]{mJ/cm^2}$) for an n-doped ($E_F$=\unit[400]{meV}) graphene sample on H-SiC (dielectric constant of $\varepsilon_s=7.8$) and on 6$\sqrt{3}$ C-SiC (dielectric constant of $\varepsilon_s=43$). The continuous, dashed, and dotted lines represent the full dynamics, the dynamics with c-ph scattering only, and the dynamics with c-c scattering only, respectively. Note that the two dashed lines overlap.}
\label{fig_screening}
\end{figure}

In Fig. \ref{fig_screening} we investigate the influence of the substrate on the amplitude and decay of the anisotropy of the photo-excited carrier distribution. The presence of a substrate can efficiently screen the Coulomb interaction by a factor of $1/\varepsilon_{\text{eff}}$, where $\varepsilon_{\text{eff}}=(1+\varepsilon_s)/2$ is related to the dielectric constant of the substrate $\varepsilon_s$ \cite{WalterPRB2011}. We investigate the two substrates used in the tr-ARPES experiment:  hydrogen-terminated SiC(0001) with a dielectric constant of $\varepsilon_s=7.8$ (green in Fig. \ref{fig_screening}) and SiC(0001) covered by a 6$\sqrt{3}$ carbon buffer layer with a dielectric constant of $\varepsilon_s=43$ (blue in Fig. \ref{fig_screening}) \cite{WalterPRB2011}. We consider n-doped samples with $E_F$=\unit[400]{meV} in the strong excitation regime ($\unit[1.5]{mJ/cm^2}$). We find a considerably slower decay of the anisotropy for graphene on 6$\sqrt{3}$ C-SiC exhibiting a large dielectric constant. This can be clearly ascribed to the drastically slower Coulomb-induced c-c scattering (see dotted lines in Fig. \ref{fig_screening}). In this situation, c-ph scattering becomes the dominant channel for redistributing the optically excited carriers and reducing the initial carrier anisotropy. Note that, in the framework of the present model, the c-ph coupling is unscreened.

\subsection{Modelling of the measured anisotropy}

\begin{figure}[t!]
 \includegraphics[width=\linewidth]{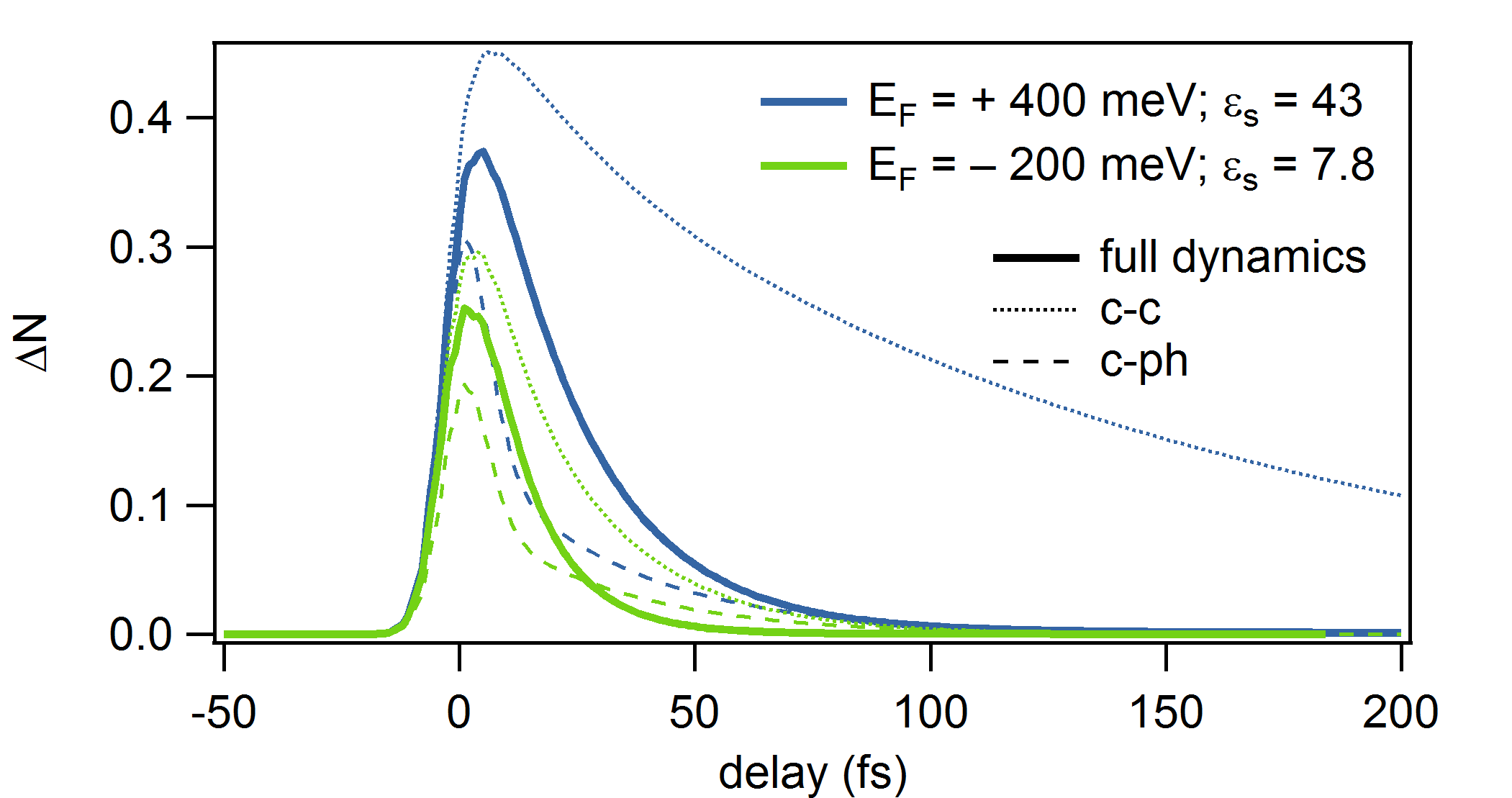}
\caption{\textbf{tr-ARPES experiment.} Temporal evolution of the carrier anisotropy $\Delta N$ at the experimental conditions from the main text: n-doped ($E_F$=\unit[400]{meV}) graphene sample on 6$\sqrt{3}$ C-SiC with a dielectric constant of $\varepsilon_s=43$ (blue) in direct comparison to a p-doped ($E_F$=\unit[-200]{meV}) sample on H-SiC with a dielectric constant of $\varepsilon_s=7.8$ (green) in the strong excitation regime ($\unit[1.5]{mJ/cm^2}$). The continuous, dashed, and dotted lines represent the full dynamics, the dynamics with c-ph scattering only, and the dynamics with c-c scattering only, respectively.}  
\label{fig_arpes}
\end{figure}

\begin{figure}
 \includegraphics[width=\linewidth]{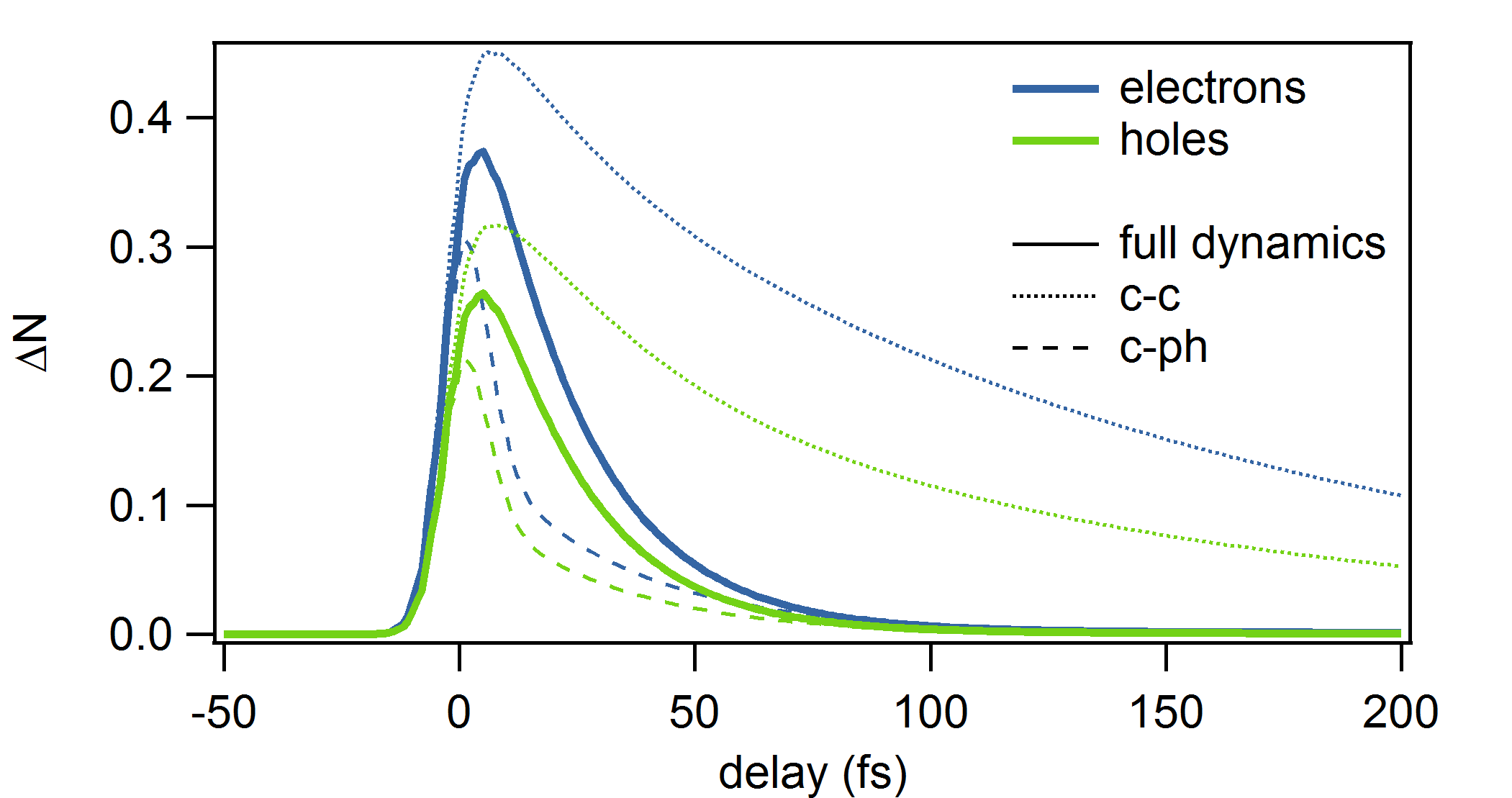}
\caption{\textbf{Electron vs hole anisotropy.} Temporal evolution of the anisotropy $\Delta N$ for electrons (blue) and holes (green) under the experimental conditions from Fig. \ref{fig_arpes}. The continuous, dashed, and dotted lines represent the full dynamics, the dynamics with c-ph scattering only, and the dynamics with c-c scattering only, respectively.}
\label{fig_e_h}
\end{figure}

Having understood the impact of pump fluence, doping, and substrate sceening, we now investigate the experimental conditions for the tr-ARPES measurements in the main part: an n-doped graphene sample on 6$\sqrt{3}$ C-SiC ($E_F$=\unit[400]{meV}, $\varepsilon_s=43$, blue in Fig. \ref{fig_arpes}) and a p-doped graphene sample on  H-SiC ($E_F$=-\unit[200]{meV}, $\varepsilon_s=7.8$, green in Fig. \ref{fig_arpes}) both excited with a strong pump fluence of $\unit[1.5]{mJ/cm^2}$. In agreement with the experimental observations, we find a more pronounced and a longer-lived carrier anisotropy for the n-doped sample compared to the p-doped sample. This can be understood by a combination of doping (Fig. \ref{fig_doping}) and screening (Fig. \ref{fig_screening}): The higher the chemical potential $\mu_e$, the smaller the scattering phase space for c-c and c-ph scattering. The larger the screening $\varepsilon_s$ the slower the c-c scattering. Due to the large value of $\varepsilon_s$ in the n-doped sample, the decay of the anisotropy is still dominated by c-ph scattering, even in the high fluence regime.

Finally, we separately study the temporal evolution of the electron anisotropy in the conduction band and the hole anisotropy in the valence band (Fig. \ref{fig_e_h}). Due to the non-zero doping of our samples, the symmetry between electron and hole dynamics is broken. In good qualitative agreement with the experimental results, we find that the hole anisotropy is smaller and decays faster. We attribute this to a smaller scattering phase space for c-c and c-ph scattering for the electrons in the conduction band compared to the holes in the valence band.


\end{document}